\documentclass[twocolumn]{aastex62}

\usepackage{color}

\shorttitle{LMC globular clusters}
\shortauthors{Piatti and Koch}

\begin{document}

\title{Search for an intrinsic metallicity spread in old globular clusters of the Large
 Magellanic Cloud}

\author[0000-0002-8679-0589]{Andr\'es E. Piatti}
\affiliation{Consejo Nacional de Investigaciones Cient\'{\i}ficas y T\'ecnicas, Av. Rivadavia 1917, 
C1033AAJ, Buenos Aires, Argentina}
\affiliation{Observatorio Astron\'omico de C\'ordoba, Laprida 854, 5000, 
C\'ordoba, Argentina}
\correspondingauthor{Andr\'es E. Piatti}
\email{e-mail: andres@oac.unc.edu.ar}

\author{Andreas Koch}
\affiliation{Zentrum f\"ur Astronomie der Universit\"at Heidelberg, Astronomisches Rechen-Institut, 
M\"onchhofstr. 12, 69120 Heidelberg, Germany}

\begin{abstract}
We report for the first time on the magnitude of the intrinsic [Fe/H] spread 
among ten old globular clusters (GCs) of the Large Magellanic Cloud (LMC).  Such spreads 
are merely observed in approximately five per cent of the Milky Way GCs and recently 
gained more attention  
in theoretical models of GC evolution. 
We derived metallicities with a typical precision of 0.05 dex $\le \sigma{\rm [Fe/H]} \le $ 0.20 dex for an average of 14 red giant branch
stars per GC from Str\"omgren photometry. The respective, metallicity-sensitive
indices have been calibrated to precise and accurate high-dispersion spectroscopy. 
For all clusters we found null [Fe/H] spreads with a typical 
uncertainty of 0.04 dex, with the possible exception of NGC\,1786 that shows an intrinsic dispersion of 0.07$\pm$0.04 dex. 
The mean, observed standard deviation of the derived metallicities for nearly 40 per cent of our GC sample amounted to
 smaller than 0.05 dex. 
At present, we cannot exclude that the remaining GCs also 
have intrinsic Fe-abundance variations in excess of 0.05 dex, but in order
to significantly detect those, the measurement errors on individual [Fe/H]-values
would need to be lowered to the 0.03--0.07 dex level.
These findings suggest, along with those from ages and light-element abundances,
that the LMC GCs studied here are alike to the majority of Galactic GCs. 
 
\end{abstract}


\section{Introduction}

Only 8 out of 156  Galactic globular clusters (GCs) 
listed in the \citep[][2010 version]{harris1996} catalog have been recently
classified as `anomalous` objects, because their intrinsic [Fe/H] dispersions are 
$>$ 0.05 dex \citep{johnsonetal2015,marinoetal2015}. From this perspective, anomalies in the iron content would not appear 
to be the most frequent manifestation of the GC multiple populations (MPs) 
formation, as recently reviewed by \citet{bl2017}. 

Among the 15 known Large Magellanic Cloud (LMC) old GCs, only NGC\,1754, 2210, and 2257 
have been searched for anomalies in their metallicities by \citet{mucciarellietal2009},
who concluded from [Fe/H] values of 5--7 stars per GC that they exhibit quite homogeneous iron abundances
(intrinsic spread=0.02--0.04 dex) despite the observed occurrence of light element variations such as an Na-O anti-correlation. Therefore, 
they show very similar properties to the vast majority of Galactic GCs.

From a theoretical point of view, different models have recently proposed distinct scenarios
to describe abundance anomalies in a variety of chemical elements in massive
clusters harboring MPs. For instance, \citet{bt2016}'s model is based on merger events,
\citet{bekki2017} and \citet{kl2018} used supernovae enrichment and asymptotic
giant branch (AGB) star ejecta, while \citet{gielesetal2018} proposed a concurrent formation of
GCs and supermassive stars, among others.
These models have been mainly stimulated by observational findings of (anti-)correlations between chemical abundances of certain light elements 
\citep[e.g., Na-O, Mg-Al, Mg-Si, Si-Zn;][]{o1971,c1978,carrettaetal2009,grattonetal2012,hankeetal2017}
and  bimodalities in CN and CH that trace light element variations \citep[e.g.][and references therein]{kayseretal2008,mg2010}. 
Some of the models also propose mechanisms to obtain
intrinsic [Fe/H] spreads $>$ 0.05 dex. For instance, \citet{gavagninetal2016}'s model used
merger events, while that of \citet{bailin2018} is based on feedback from core-collapse supernovae.
Recently, \citet{limetal2017} found that GCs with large intrinsic [Fe/H] spreads
also show a positive CN-CH correlation.

In this work we report on the magnitude of the intrinsic [Fe/H] spread for a sample of 
nine old LMC GCs and in ESO\,121-SC03, which lies in the LMC's GC age gap \citep[9 Gyr,][]{betal98,mackey2006}.  As far as we are aware, this is the 
largest sample of LMC GCs -- assumed to have MPs but not confirmed as yet -- analyzed in order to 
search for internal metallicity variations. In Section 2 we describe the observational material and precision and 
accuracy of our photometry. Section 3 deals with the estimation of individual metallicities for
carefully selected cluster red giant branch (RGB) stars, while in Section 4 we analyze and discuss
our results. 

\section{Observational data sets}

We made use of publicly available images of ten LMC GCs obtained through the 
Str\"omgren $vby$ filters with the SOAR Optical Imager (SOI) mounted on the 4.1-m
Southern Astrophysical Research (SOAR) telescope (program SO2008B-0917,
PI: G. Pietrzy\'nski). The images (field of view = 5.25$\arcmin$$\times$5.25$\arcmin$, 
pixel scale = 0.154$\arcsec$/px) 
were obtained from exposures of 350--500, 140--300 and 90--180 s in the
$v$, $b$, and $y$ filters, respectively, under excellent atmospheric conditions
(FWHM $\sim$ 0.6$\arcsec$). We processed the images, along with the
respective calibration images, following the SOI 
pipeline\footnote{http://www.ctio.noao.edu/soar/content/soar-optical-imager-soi}.

Stellar photometry on the individual images was obtained using the
{\sc daophot} package \citep{setal90}. We first derived a quadratically varying point-spread-function 
(PSF) by using two sets of stars selected interactively, one with $\sim$100 and another 
with the brightest $\sim$40 stars. The smaller group was used to build a preliminary
PSF, which was in turn used to clean the larger PSF sample. The resulting PSF
was applied to all the identified stars in an image, which were then subtracted from it 
in order to identify fainter stars, and to run again {\sc allstar} on the enlarged 
star sample. We iterated the loop three times. 
Finally, we kept only sources
with $\chi$ $<$ 2, {\sc $|$sharpness$|$} $<$ 0.5, and  {\sc daophot} photometric errors smaller than  0.01 mag.

To standardize the PSF photometry we first measured instrumental magnitudes of 5--10 standard stars
observed twice per night, thus covering an airmass range between 1.1 and 2.1.
Then, we fitted the transformation  equations given by the expressions:
\begin{eqnarray*}
v &\,=\,& v_1 + V_{\rm std} + v_2\times X_v + v_3\times (b-y)_{\rm std} + v_4\times m_{\rm 1 std},\\
b &\,=\,& b_1 + V_{\rm std} + b_2\times X_b + b_3\times (b-y)_{\rm std},\\
y &\,=\,& y_1 + V_{\rm std}  + y_2\times X_y + y_3\times (b-y)_{\rm std},
\end{eqnarray*}
where  $v_i$, $b_i$ and $y_i$ are the i-th fitted 
coefficients, and $X$ represents the effective airmass. Table~\ref{tab:table1} shows
the resulting coefficients obtained using the {\sc iraf.fitparams} routine.

\begin{deluxetable*}{ccccccc}
\tablecaption{Str\"omgren transformation coefficients.\label{tab:table1}}
\tablehead{\colhead{Filter} & \colhead{coef$_{\rm 1}$} & \colhead{coef$_{\rm 2}$} & \colhead{coef$_{\rm 3}$} & \colhead{coef$_{\rm 4}$} & \colhead{rms}}
\startdata
\multicolumn{6}{l}{Dec. 18, 2008: NGC\,2257, ESO\,121-SC3} \\\hline
$v$    &  1.122$\pm$0.007& 0.295$\pm$0.005& 1.995$\pm$0.048& 1.026$\pm$0.061 &  0.002 \\
$b$    & 0.942$\pm$0.014& 0.177$\pm$0.009 &0.946$\pm$0.014  &    $\ldots$   &0.008        \\ 
$y$    & 0.932$\pm$0.015    &0.122$\pm$0.009 &-0.005$\pm$0.016  &    $\ldots$   &0.010        \\ 
\hline
\multicolumn{6}{l}{Dec. 19, 2008: NGC\,1754, 1786, 1835, 1898, 2005, 2019, 2210} \\\hline
$v$    &  1.096$\pm$0.015& 0.286$\pm$0.009& 2.004$\pm$0.030& 1.117$\pm$0.038 &  0.010 \\
$b$    & 0.916$\pm$0.013& 0.169$\pm$0.007 &0.999$\pm$0.011  &    $\ldots$   &0.010        \\ 
$y$    & 0.939$\pm$0.019    &0.107$\pm$0.010 &0.018$\pm$0.015  &   $\ldots$    &0.016        \\ 
\hline
\multicolumn{6}{l}{Jan. 16, 2009: NGC\,1841} \\\hline
$v$    &  1.005$\pm$0.004& 0.290$\pm$0.009& 2.034$\pm$0.032& 0.914$\pm$0.028 &  0.007 \\
$b$    & 1.014$\pm$0.007& 0.170$\pm$0.007 &0.939$\pm$0.018  &   $\ldots$    &0.011        \\ 
$y$    & 1.005$\pm$0.004    &0.120$\pm$0.010 &-0.046$\pm$0.011  &    $\ldots$   &0.007        \\ 
\enddata
\end{deluxetable*}

The quality of our photometry was first examined in order to obtain robust estimates of the photometric 
errors. To do this, we performed artificial star tests by using the stand-alone {\sc addstar} program 
in the {\sc daophot} package \citep{setal90} to add synthetic stars, 
generated bearing in mind the color and magnitude distributions 
of the stars in the color-magnitude diagram (CMD)  as well as the cluster radial stellar 
density profile. We added a number
of stars equivalent to $\sim$ 5$\%$ of the measured stars in order to avoid in the synthetic images 
significantly 
more crowding than in the original images. On the other hand, to avoid small number statistics in the
 artificial-star 
analysis, 
we created a thousand different images for each original one. We used the option of entering the number of
 photons
per ADU in order to properly add the Poisson noise to the star images. 

\begin{figure*}[htb]
     \includegraphics[width=\columnwidth]{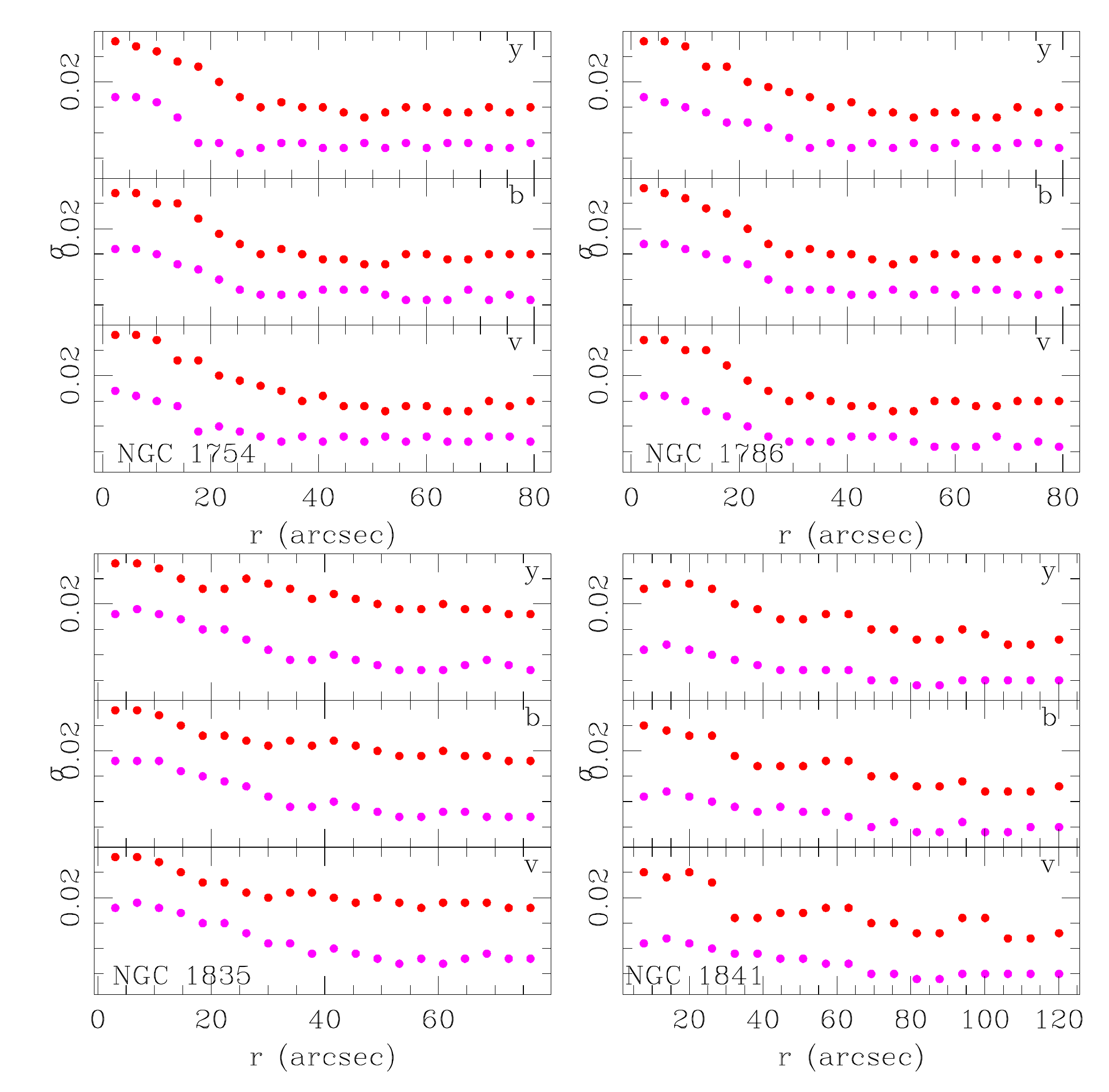}
     \includegraphics[width=\columnwidth]{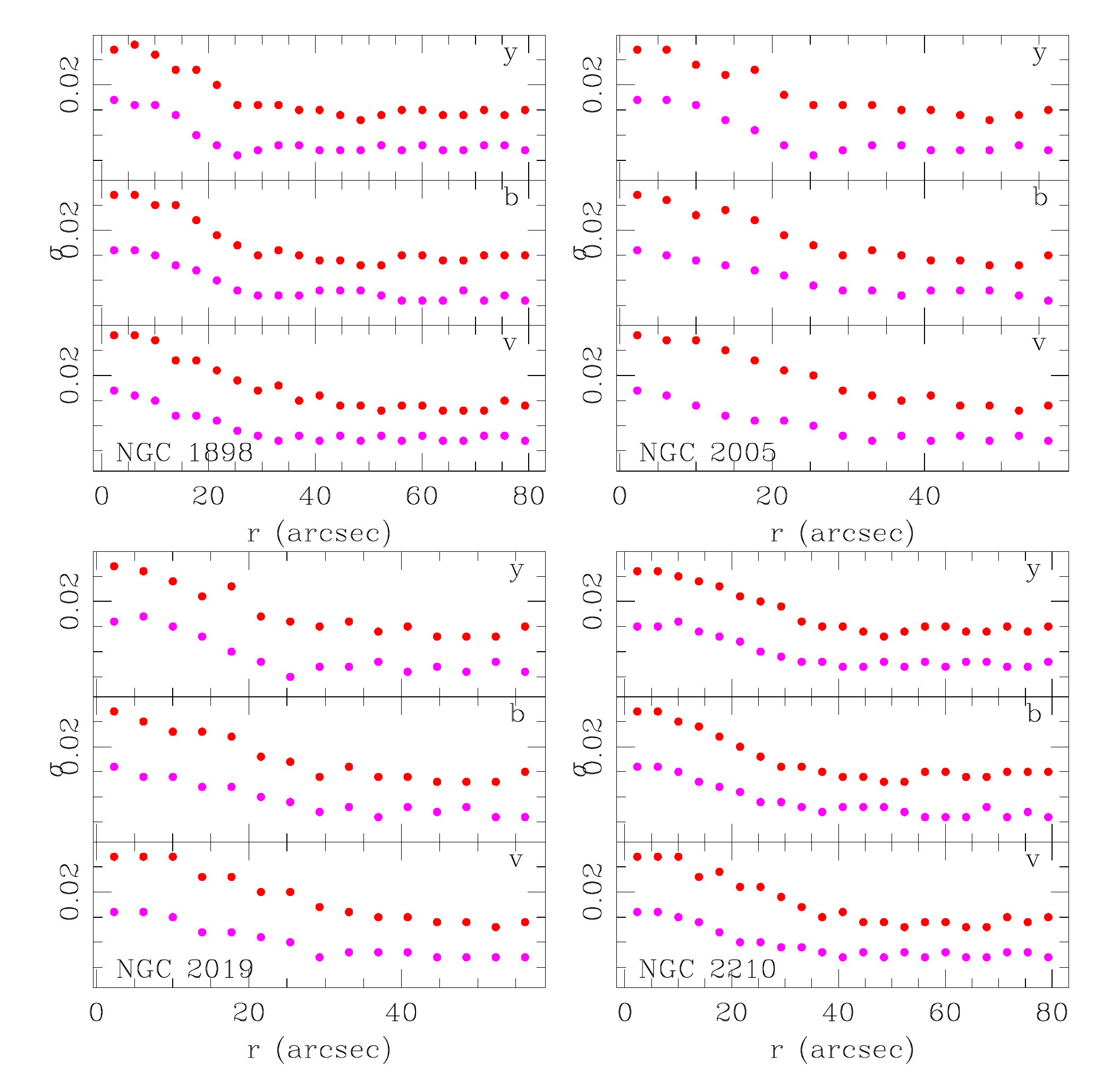}
     \includegraphics[width=\columnwidth]{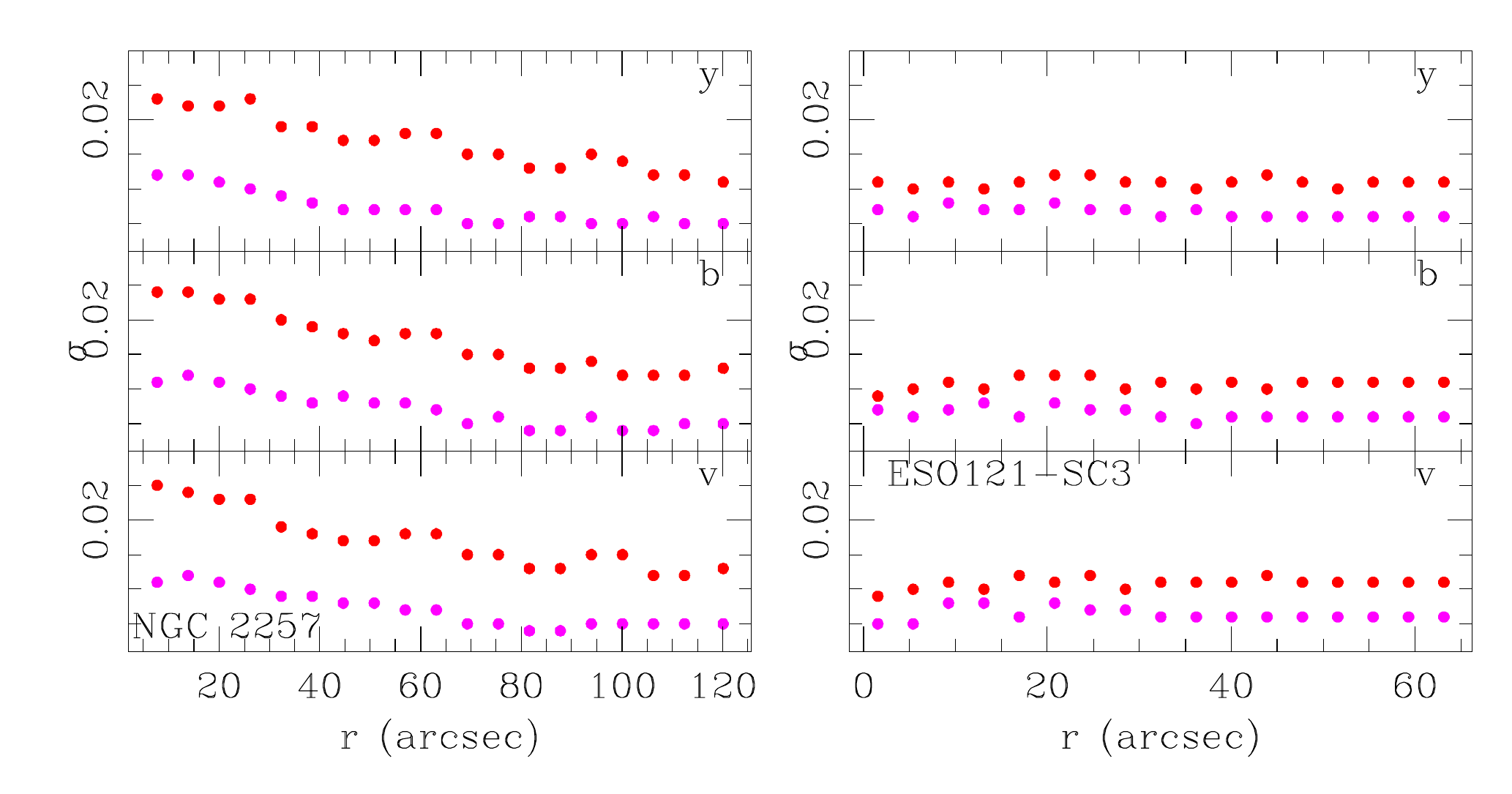}
   \caption{Photometric error estimates for the  $vby$ filters as a function of the distance
to the cluster center. Magenta and red filled circles are for $V$ = 16.5 and 18.5 mag, respectively. }
 \label{fig1}
\end{figure*}

We then repeated the same steps to obtain the photometry of the synthetic images as described above, 
i.e., 
performing three passes with the {\sc daophot/allstar} routines. 
The photometric errors were derived from the magnitude difference between the output and input data
of the added synthetic stars using the {\sc daomatch} and {\sc daomaster} tasks. We found that this difference
resulted typically equal to zero and in all the cases smaller than 0.003 mag. The respective rms errors were 
adopted as the photometric errors.
Fig.~\ref{fig1} illustrates the behavior of these errors as a function of the distance from the cluster center and of the magnitude. For clarity, we only show
two different magnitude level, namely $V$ = 16.5 mag and 18.5 mag, respectively. 
These magnitudes
roughly correspond to the upper and lower limits of the cluster RGBs used in this work
(see Fig.~\ref{fig2}). 

\section{Metallicity estimates}

We used the $m_{\rm 1}$ vs. $b-y$, $m_{\rm 1}$ vs. $v-y$, $V$ vs. $b-y$, and 
 $V$ vs. $v-y$ diagrams as discriminators between giant, subgiant,
and dwarf stars (thus weeding out foreground stars) to select  RGB stars in stellar populations
\citep{faria2007,arnadottir2010,calamidaetal2012,calamidaetal2014}. We contrained our search to
RGB stars located inside the cluster radii estimated by \citet{pm2018}, and brighter that 
the respective cluster horizontal branch in the $V$ vs.
$b-y$ CMD, where the RGBs are narrower and least contaminated
by field stars \citep{franketal2015}. 
 Indeed, we only kept stars within a strip of $\pm$0.05 mag in $b-y$ from the cluster RGB ridge lines, assumed the $b-y$ color
to be mainly a temperature effective indicator with less metallicity sensitivity \citep{cm1976}. 
We also discarded any star lying inside the
RGB strips whose [Fe/H] value departs significantly (more than 3 times the observed dispersion)
from the readily visible cluster
metallicity distributions. In addition,
by using different field regions of equal cluster areas, distributed around the clusters, we
found that the number of field stars which fall inside the RGB strips is smaller
than 10 per cent with respect to the total number of adopted cluster members.
Fig.~\ref{fig2} shows all the measured stars inside the cluster radius, 
those located far from the cluster region for an equal cluster area, and the selected
RGB stars drawn with black,  orange and red filled circles, respectively.

By using the  semi-empirical calibration by \citet{calamidaetal2007}, we estimated individual
metallicities ([Fe/H]) from the reddening corrected metallicity index\footnote{Using the standard 
definition $m_1  = (v-b) - (b-y)$.}
 $m_{\rm 1}$$_{0}$,
in turn based on the stars' $(v-y)_0$ colors.
The uncertaintes were calculated from propagation of all of the involved errors, i.e., those of the
calibration and those of our photometry,  as follows:

\begin{displaymath}
{m_{\rm 1}}_0 = \alpha + \beta {\rm [Fe/H]} + \gamma (v-y)_0 + \delta {\rm [Fe/H]}(v-y)_0, 
\end{displaymath}

where $\alpha$ = $-$0.309, $\beta$=$-$0.090$\pm$0.002, $\gamma$=0.521$\pm$0.001, and
$\delta$=0.159$\pm$0.001, respectively. Then, per standard error propagation, 
\begin{eqnarray*}
\sigma({\rm [Fe/H]})^2 &\, = \, & \left(\frac{\partial {\rm [Fe/H]}}{\partial \alpha} \sigma(\alpha)\right)^2 
      + \left(\frac{\partial {\rm [Fe/H]}}{\partial \beta} \sigma(\beta)\right)^2 \\
      &\,+\,&
   \left(\frac{\partial {\rm [Fe/H]}}{\partial \gamma}\sigma(\gamma)\right)^2 + 
     \left(\frac{\partial {\rm [Fe/H]}}{\partial \delta}\sigma(\delta)\right)^2 \\
     &\,+\,& 
    \left(\frac{\partial {\rm [Fe/H]}}{\partial {m_{\rm 1}}_0}\sigma({m_{\rm 1}}_0)\right)^2 + 
  \left(\frac{\partial {\rm [Fe/H]}}{\partial (v-y)_0}\sigma((v-y)_0)\right)^2\\
&\,=\,& \left(\frac{0.002{\rm [Fe/H]}}{c}\right)^2 + 
\left(\frac{0.001(v-y)_0}{c}\right)^2 \\
&\,+ \,&\left(\frac{0.001{\rm [Fe/H]}(v-y)_0}{c}\right)^2 +
\left(\frac{\sigma({m_{\rm 1}}_0)}{c}\right)^2 
\end{eqnarray*}
\vspace{-2ex}
\begin{displaymath}
+  \left(\frac{(-0.521c - 0.159({m_{\rm 1}}_0 +0.309 -0.521(v-y)_0)\sigma((v-y)_0)}{c^2}\right)^2, 
\end{displaymath}
 where c = $-0.090 + 0.159(v-y)_0$, and $\sigma({m_{\rm 1}}_0)$ and $\sigma((v-y)_0)$
are the photometric errors in ${m_{\rm 1}}_0$ and $(v-y)_0$, respectively, according to
the position of the stars with respect to the cluster's center (see Fig.~\ref{fig1}). 
We note that ${m_{\rm 1}}_0$ and $(v-y)_0$ depend on the $vby$ magnitudes of each star, in the sense that the fainter a star, the larger its errors. 
They also depend on the position
of the star respect to the cluster center, because of crowding effects. Typically, stars 
located in the outer regions have smaller magnitude errors as compared with those of the
same magnitudes placed at the cluster center. We
derived them as follows:
\begin{eqnarray*}
(v-y)_0 \,&= \,&v-y - 1.67\times 0.74E(B-V),\\
\sigma((v-y)_0)^2 &\,=\,& \sigma(v)^2 + \sigma(y)^2 + (1.67\times 0.74\sigma(E(B-V)))^2 
\end{eqnarray*}
and
\begin{eqnarray*}
 {m_{\rm 1}}_0 &\,=\,& (v-b) - (b-y) + 0.33 \times 0.74E(B-V),\\
\sigma({m_{\rm 1}}_0)^2  &\,=\,& \sigma(v)^2 + 4\sigma(b)^2 + \sigma(y)^2  \\
& & \hspace{0.9cm} + (0.33\times 0.74\sigma(E(B-V)))^2,
\end{eqnarray*}
where the reddening laws,  $E(X)/E(B-V)$,  are those given by \citet{cm1976}.
We obtained the cluster reddening values from
the NASA/IPAC  Extragalactic Data base\footnote{http://ned.ipac.caltech.edu/. NED is operated 
by the Jet Propulsion Laboratory, California Institute of Technology, under contract with 
NASA.} (NED) (see Table~\ref{tab:table2}).  
The resulting [Fe/H] values and their errors are plotted
in the bottom-right panels of Fig.~\ref{fig2}.
\subsection{Intrinsic dispersions}
We  next determined the mean and dispersion of each cluster's Fe-abundance by
employing a maximum likelihood approach, similar to the method detailed in \citet{franketal2015}.
The relevance lies in accounting for each individual star's measurement errors, which could 
artificially inflate the dispersion if ignored. 
We optimized the probability $\mathcal{L}$ that a
given ensemble of stars with 
metallicities [Fe/H]$_i$ and metallicity errors $\sigma_i$ are drawn from a population with mean
Fe-abundance $<$[Fe/H]$>$ and intrinsic dispersion W  
\citep[e.g.,][]{Walker2006,Koch2018Gaia}, 
as follows:
\begin{eqnarray*}
\mathcal{L}\,=\,&\prod_{i=1}^N&\, \left(\,2\pi\left[\sigma_i^2 + W^2 \, \right]\,\right)^{-\frac{1}{2}} \\
&\times&\,\exp \left(-\frac{\left([{\rm Fe/H}]_i \,- <[{\rm Fe/H}]>\right)^2}{\sigma_i^2 + W^2} \right), 
\end{eqnarray*}
where the errors on the mean and dispersion were computed from the respective covariance matrices. 
 We note that this approach assumes that the error distribution is Gaussian, which is adopted here because of the limited number of stars
\citep[cf.][]{franketal2015}.
The last columns of Table~\ref{tab:table2} list the mean metallicity,  the intrinsic 
dispersion, and the number of stars used for each cluster. 
As far as we are aware, these comprise the largest sample of LMC GCs with metallicities put 
on an homogeneous scale. Here, we note the remarkably tight correlation between the spectroscopic 
and Str\"omgren-based metallicities, attesting to the validity of the calibrations over a broad 
metallicity- and color-range (see Fig.~\ref{fig3}).  Moreover, although the CN absorption band at
$\lambda$4142\AA~is near to the effective wavelength of the $v$ filter, our metallicities
appear to be  also driven by Fe abundances, as the index $m_1$ was calibrated 
by \citet{calamidaetal2007} as a photometric proxy for the iron abundance. Nevertheless,
we cannot rule out that they could also reflect CN variations \citep{limetal2017}.
\subsection{ Foreground contamination}
 Our measured intrinsic dispersions values are not expected to be affected 
by field star contamination, since the LMC field stellar population have more metal-rich mean 
metallicities
 \citep[e.g.,][]{cole2000,pompeia2008,carrera2011,vds13,pg13}, and Milky Way (MW) stars placed along the cluster 
RGB sequences are expected to be negligible across the relative small cluster areas
 \citep[$r$$\la$1.5$\arcmin$][]{pm2018}.

We show in the $\sigma$[Fe/H] vs. [Fe/H] plots
of Fig.~\ref{fig2} every measured field star located in the observed cluster areas, and
distributed outside the adopted cluster RGB strips (black circles). For comparison 
we also included field stars from regions outside the cluster areas (orange circles).
As can be seen, these do not visibly widen the metallicity range of the selected stars.
We note that field RGB stars younger than the GCs have [Fe/H] $>-$1.0 dex \citep{coleetal2005,pg13}
and are placed along GC RGBs with [Fe/H] $> -$1.0 dex, depending on their ages \citep{getal03,os15}.

In order to assess the influence of LMC field star contaminants on our measurements, 
we expanded the above likelihood estimator by adding a fraction $f$ of foreground 
stars \citep{koch2007}:
\begin{eqnarray*}
\mathcal{L}\,=\,&\prod_{i=1}^N& \,(1-f)\, \, \left(\, 2\pi\,\left[ (\sigma_i^2 + W^2 \, \right] \right)^{-\frac{1}{2}} \\
&\times&\,\exp \left(-\frac{([{\rm Fe/H}]_i \,- <[{\rm Fe/H}]>)^2}{\sigma_i^2 + W^2} \right), \\
&+&f\,p([{\rm Fe/H}])
\end{eqnarray*}
Here, the metallicity distribution function $p($[Fe/H]$)$ of the contaminants was 
drawn from the observed Str\"omgren-metallicity distribution of LMC field stars of \citet{cole2000}. 
This was done exemplary for NGC~1786, the only object in our sample that shows a hint of a significant intrinsic spread. 
As a result, 
even fractions as high as $>$90\% will leave the result significant at the 1$\sigma$-level.

Secondly, we removed stars from the NGC 1786 sample at random in a jackknife manner, recomputing the mean and dispersion as before. 
As a result, even upon rejection of three stars (amounting to 20\% of the entire sample) as if they were foreground contaminants, 
did not alter the measured, non-negligible intrinsic dispersion and maintained the significance of the result. 
As for the other GCs with null dispersions, their results also remained unaffected, yielding the low-to-zeros values as before.

\subsection{ AGB star contamination}

Stellar populations such as GCs in the MW and the LMC can be expected to host 
a non-zero population of AGB stars, which can add a contamination of up to 10\% of stellar samples as 
analysed here \citep[e.g.][]{Kamath2010}.  
This is exemplified in  ESO 121-SC3 (bottom panel of Fig.~\ref{fig2}), which 
has six {\em bona fide} members  in the analysis. The scatter in
metallicity appears dominated by a single object yielding a low metallicity. 
As the brightest star observed in the cluster, near the
tip of the RGB we cannot refute that this star is a potentially variable 
star on the AGB, rather than an RGB star. 

To assess their influence on the metallicity spreads, we first consulted a set of Padova isochrones
\citep{Bressan2012}, available in the St\"omgren system, with ages and metallicities tailored to the 
individual GCs of this study.
We then created random samples of stars matching the observed numbers. and designed them to lie either on the theoretical RGB or AGB tracks; 
next, we computed their metallicities from the above calibrations, using the isochrone's ($v-y$) and $m_1$ colors. 
Albeit the isochrones', per construction, mono-metallicity, this procedure already introduced a standard deviation on the order of 0.04 dex
from this theoretical consideration for either class of stars. 
Moreover, metallicities derived from AGB stars are on average lower by merely 0.02 dex, so that the inclusion of a minor contamination 
(1--2 stars per GC)  in our, presumed RGB samples, would not lead to a significant inflation of the intrinsic dispersion if they were misclassified as RGB stars.  

However, we note that, as being warmer and less massive than RGB stars, metallicities of AGB stars should  follow different relations of the 
\citet{calamidaetal2007} calibration that used here, also due to the different strength of the molecular bands, with weaker CN- and CH-bands.
Ideally, the AGB contamination should be weeded out by using the index $c_1=(u-v)-(v-b)$ that is sensitive to stellar evolutionary stage \citep{arnadottir2010},
but this is inhibited by the lack of any $u$-band observations for our LMC targets.
\section{Analysis and discussion}
 The resulting dispersions reveal that nine of the ten LMC GCs studied here
show metallicity spreads consistent with zero within the respective errors, as it is the case for most of 
the 
MW GCs. 
One possible exception is NGC 1786, at $W=0.07\pm0.04$ dex. 
The posteriori likelihood distribution for its mean metallicity and intrinsic dispersion is shown in 
Fig.~\ref{fig4}.
We note that all our results  include a  thorough consideration of all errors 
propagated through the calibration for each star, giving
a realistic account of the, non-necessarily symmetric, error distribution.

 In order to obtain W values larger than 0.05 dex, errors
of the individual metallicities of our selected stars
would need to be smaller than 0.04 dex (NGC\,1754), 0.10 dex (NGC\,1786), 0.03 dex (NGC\,1835),
0.07 dex (NGC\,1841), 0.03 dex (NGC\,1898), and 0.06 dex (NGC\,2005), respectively. For
the remaining GCs the standard deviation of the derived [Fe/H] values is smaller than
0.05 dex, so that we discarded those GCs as possible candidates for Fe-abundance anomalies.
They represent 40 per cent of the whole analyzed sample. The estimated upper limits in the [Fe/H] 
uncertainties put a demanding constraint on the precision of [Fe/H] estimates. Indeed,
by using only NGC\,1786's stars with a tighter error constraint $\sigma$[Fe/H] $<$ 0.10 dex, we derived W = 0.07$\pm$0.04
dex (N=8). For other GCs we did not reach the expected metallicity uncertainties, which
could be attempted to be obtained from high S/N, high-dispersion spectroscopy.

As far as we are aware, there is no study showing the existence of MPs among the LMC's old GC population
from their color-magnitude diagrams \citep[cf.][]{Milone2009}, although three of them  display the 
give-away light element abundance signatures such as a Na-O anti-correlation. 
In fact, as equivalents of MW GCs, they are assumed to harbor MPs, 
and there is an acceptable synchronicity between their ages \citep{wagnerkaiseretal2017} and the abundances of some light chemical
elements \citep{mucciarellietal2010}, and overlap of their mass and age to relaxation-time ratio ranges 
\citep{pm2018}. As the [Fe/H]-spread is concerned, the eight MW GCs with Fe-abundance variations
larger than 0.05 dex represent nearly five per cent of the whole GC population. If such a small percentage
were applicable to the LMC GC population, we should find no more than one GC with such a metallicity spread, in agreement with our findings for NGC\,1786 (at a significance of $\sim 1.7\sigma$).

This work represent the first attempt in measuring the iron abundance variation in most of the LMC GCs
so far. We were motivated by the fact that Fe-abundance variations are found among some MW GCs
and by the increasing number of theoretical models that predict the presence of such chemical 
inhomogeneities. We used on average 14 selected RGB stars per cluster and obtained [Fe/H] values with 
uncertainties smaller than 0.20 dex, and nearly comparable accuracy to that derived from high-dispersion
spectroscopy for some of the brightest stars. We tightly reproduced their generally 
accepted spectroscopic metallicity scale and found no hint -- except for NGC\,1786 -- 
for a metallicity spread. This outcome supports that LMC GCs share chemical abundance patterns similar to 
those seen in many of their Galactic counterparts, as has been suggested from the
comparative, high-resolution abundance measurements in the LMC GCs \citep{mucciarellietal2010,bl2017}.

\begin{figure*}
\includegraphics[width=\columnwidth]{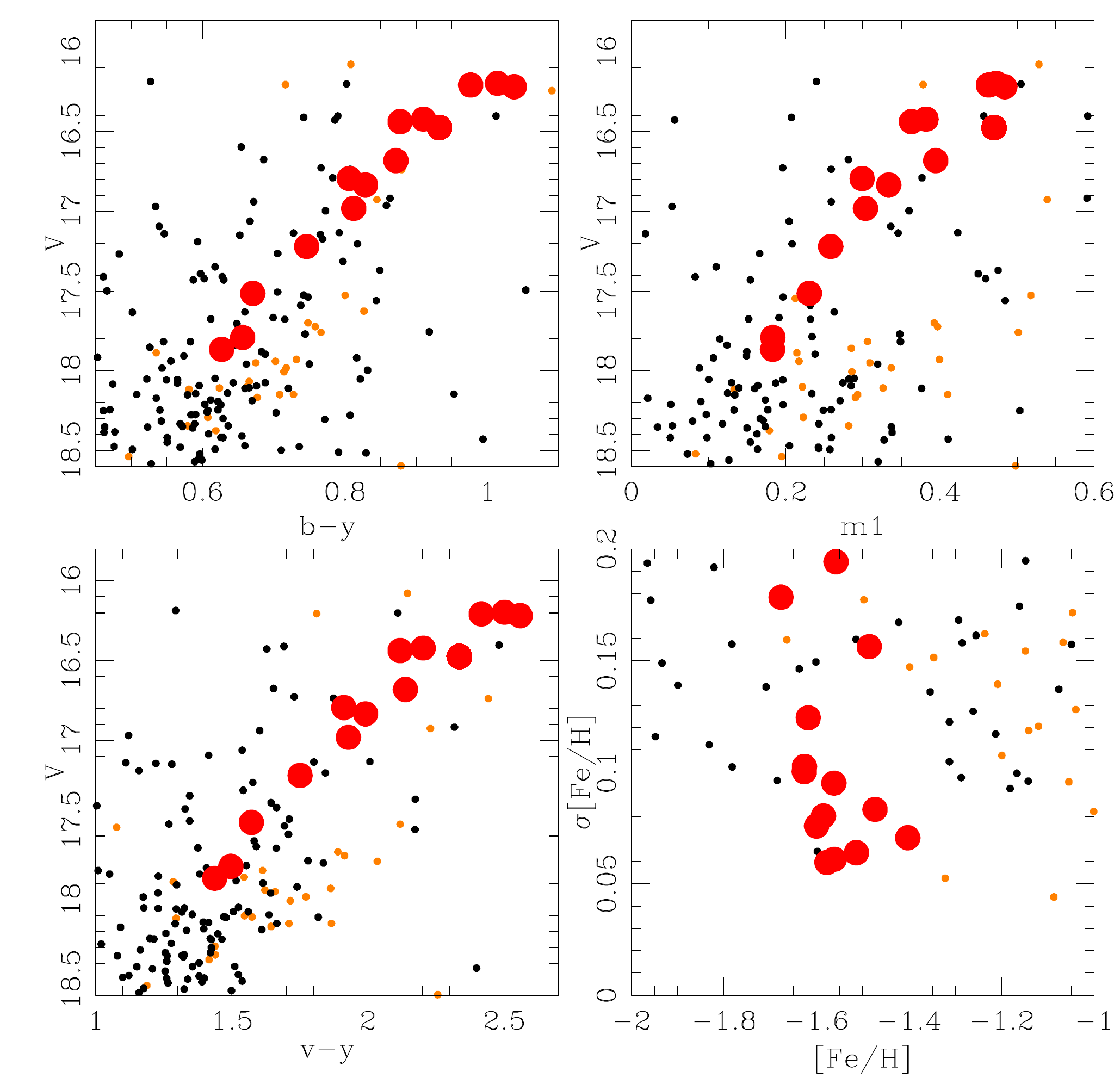}
\includegraphics[width=\columnwidth]{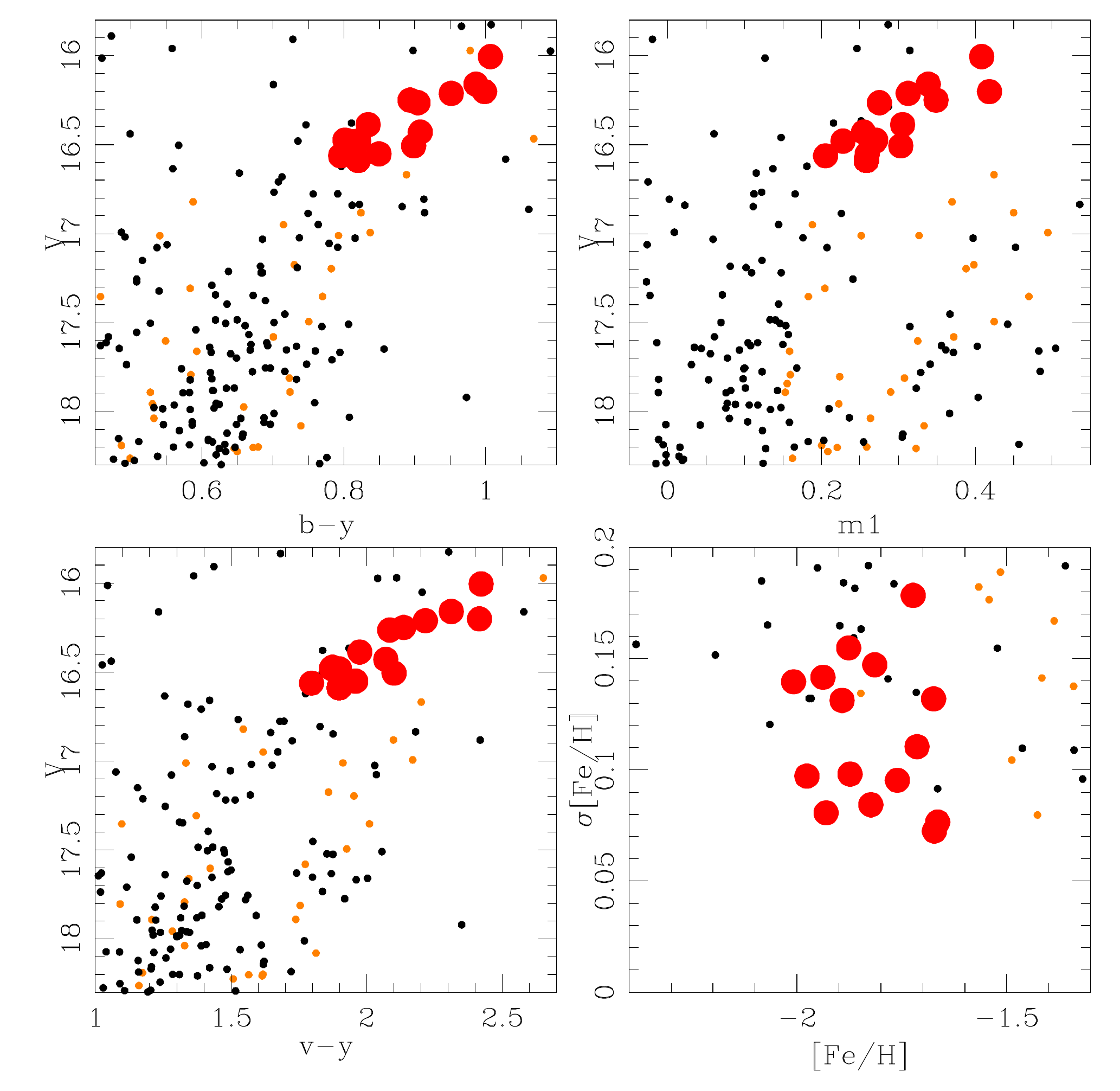}
\includegraphics[width=\columnwidth]{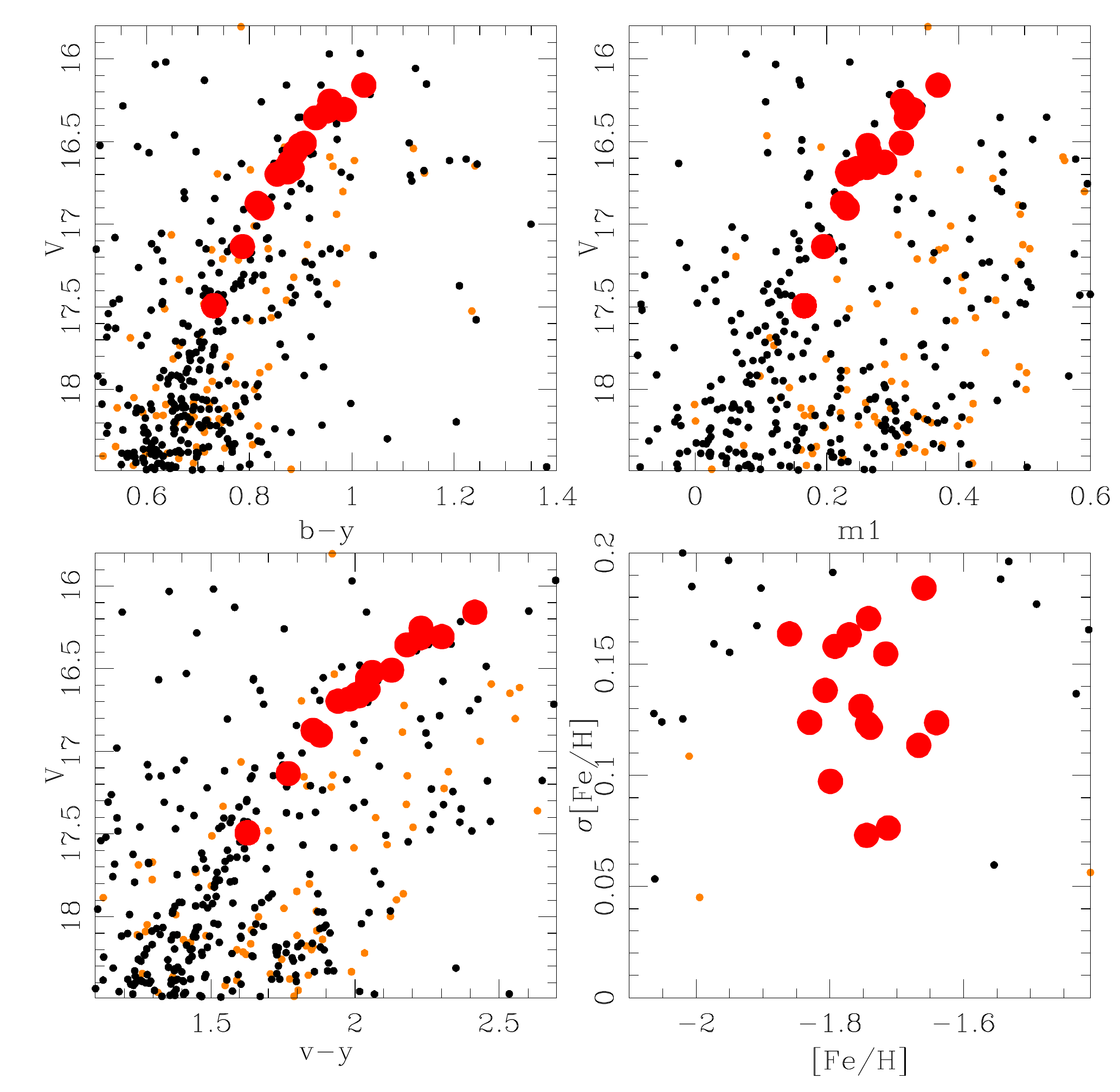}
\includegraphics[width=\columnwidth]{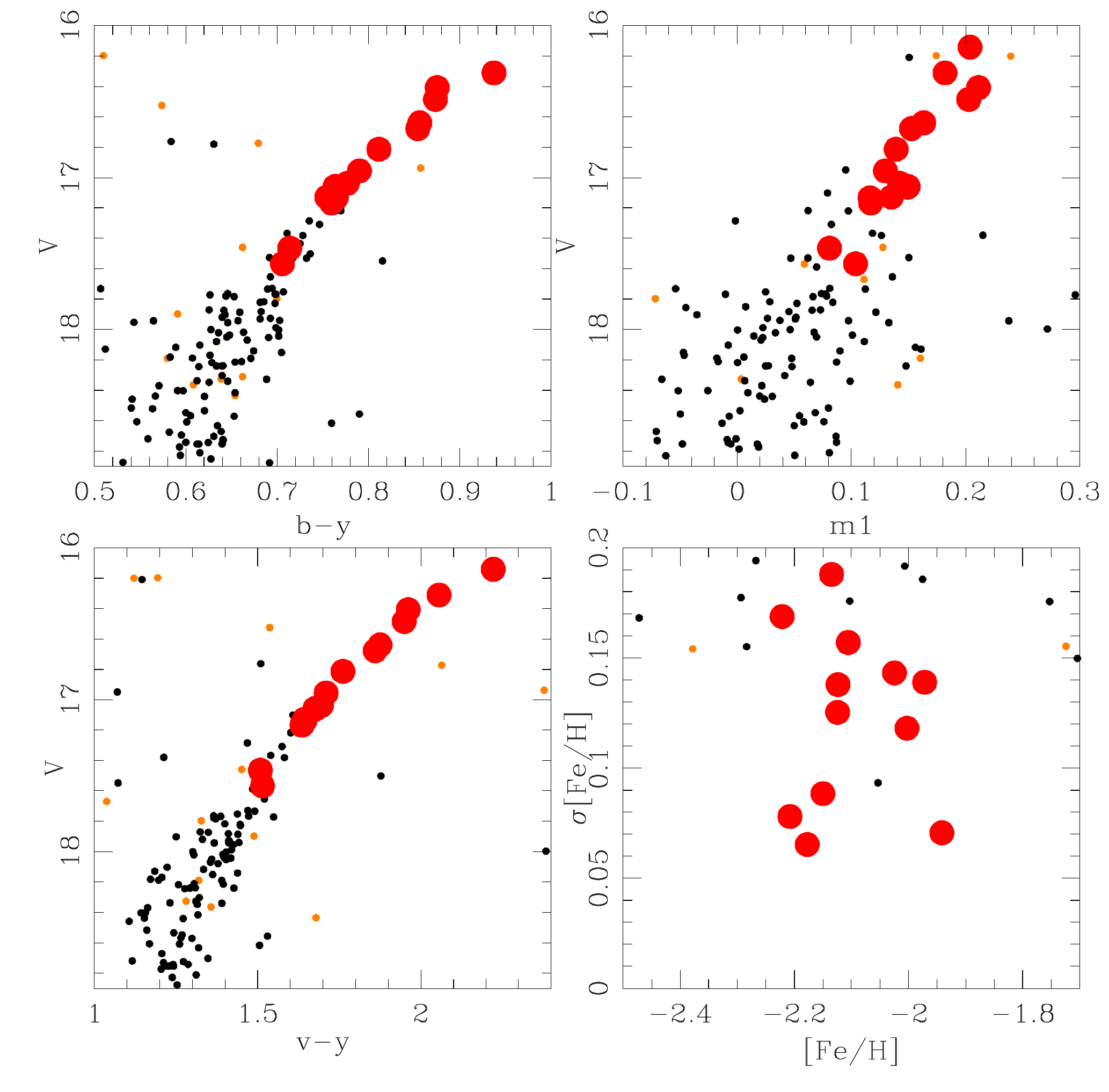}
\caption{CMDs of stars observed in the field of the GCs NGC\,1754 (top-left),
NGC\,1786 (top-right), NGC\,1835 (bottom-left), and NGC\,1841 (bottom-right). Black and
 orange filled circles represent stars observed in the cluster area and in
a comparison star field with an equal cluster area, respectively. Selected stars
are drawn with larger red circles. The individual metallicities 
([Fe/H]) and their respective uncertainties are depicted in the bottom-right panel. 
respectively. }
\label{fig2}
\end{figure*}

\setcounter{figure}{1}
\begin{figure*}
\includegraphics[width=\columnwidth]{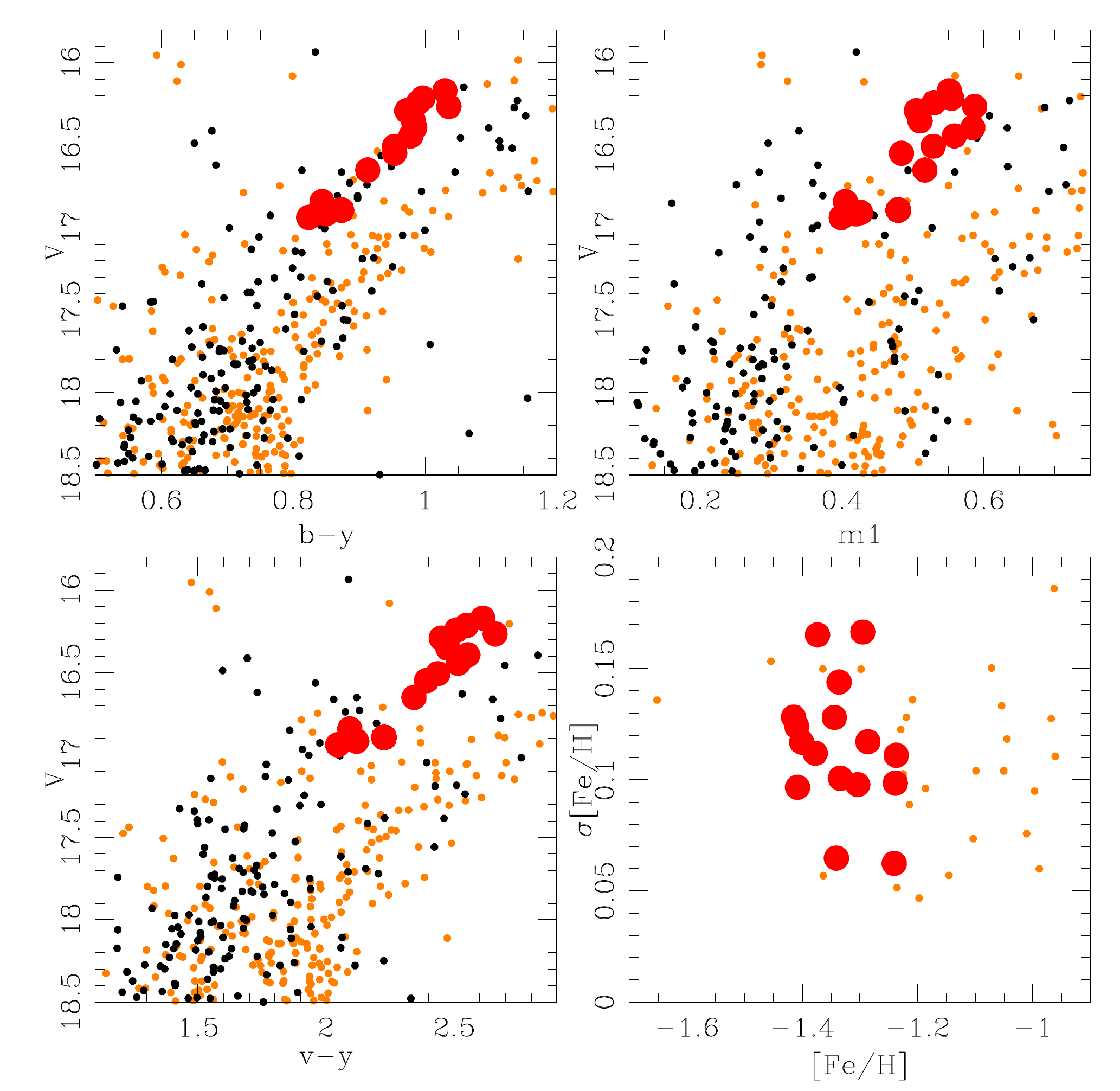}
\includegraphics[width=\columnwidth]{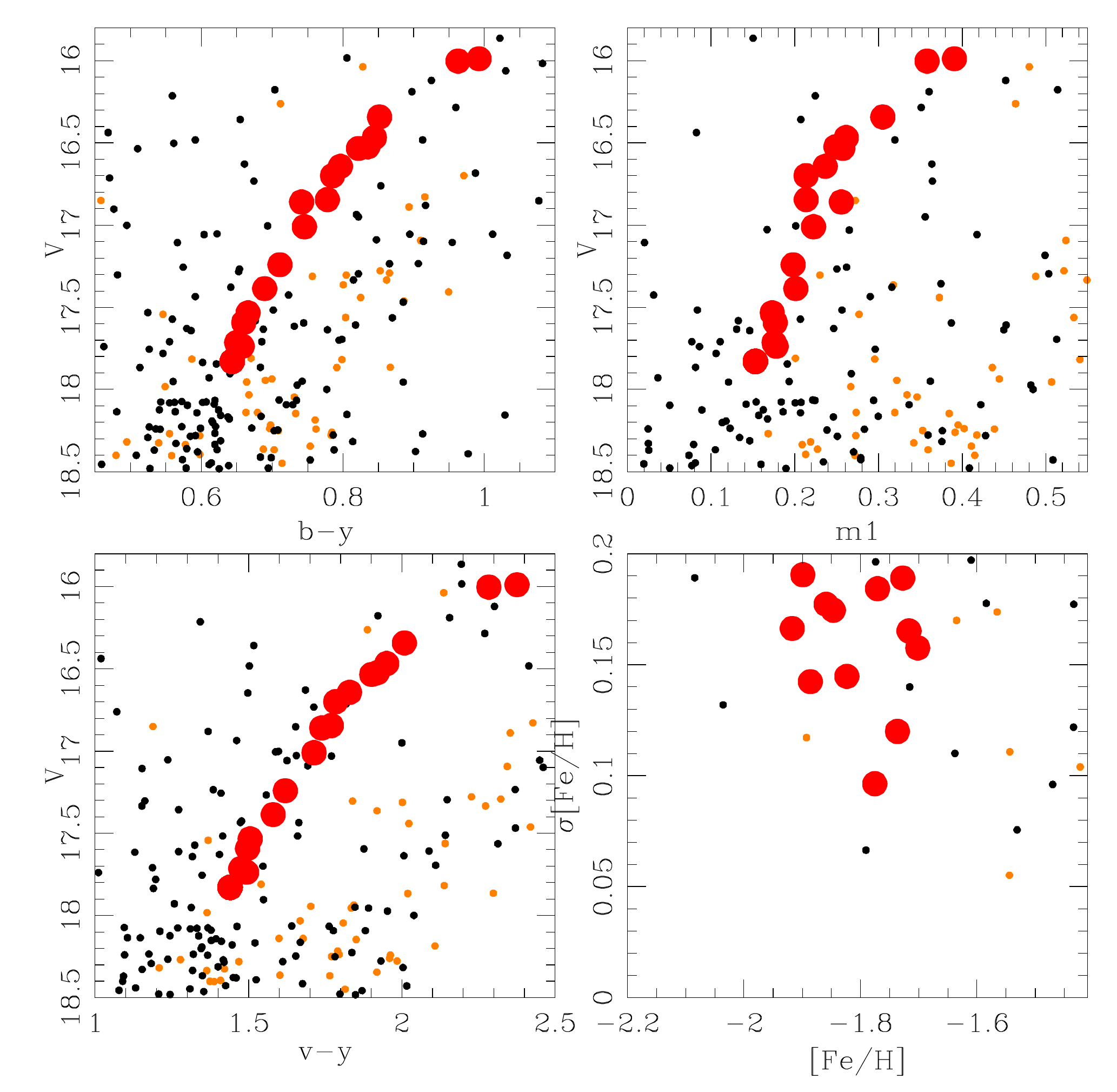}
\includegraphics[width=\columnwidth]{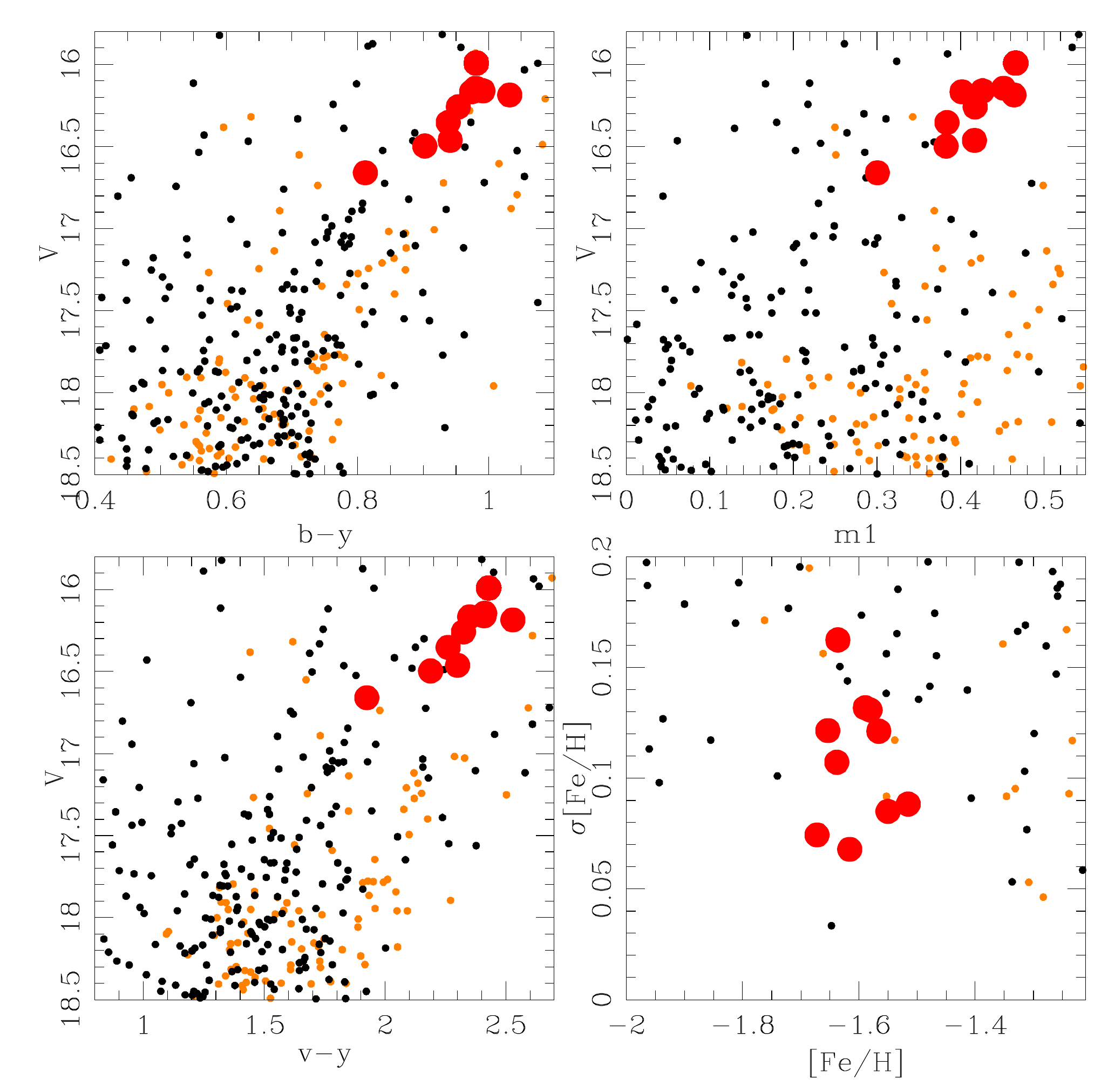}
\includegraphics[width=\columnwidth]{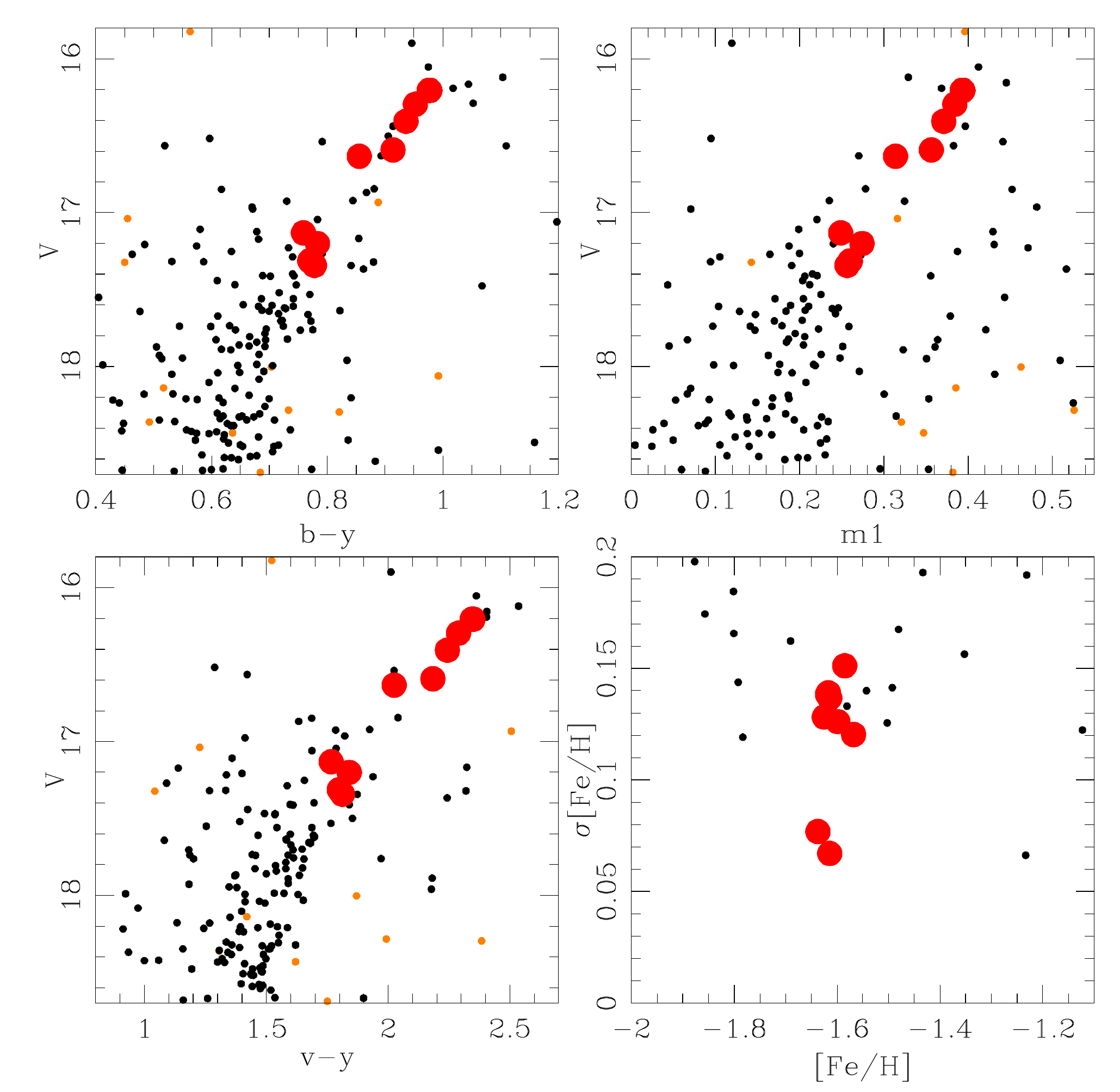}
\caption{ (continued). NGC\,1898 (top-left),
NGC\,2005 (top-right), NGC\,2019 (bottom-left), and NGC\,2210 (bottom-right).}
\end{figure*}

\setcounter{figure}{1}
\begin{figure*}
\includegraphics[width=\columnwidth]{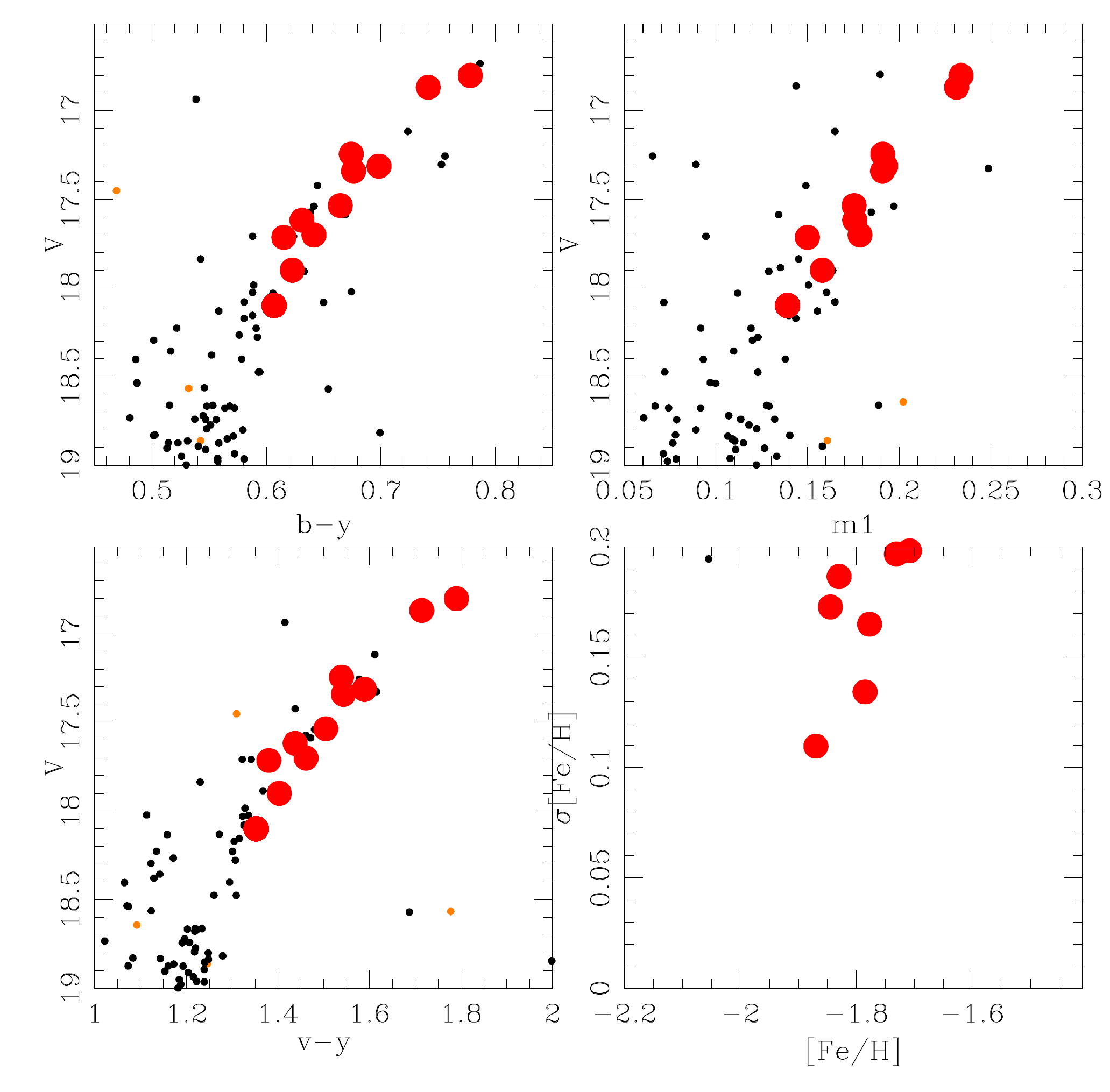}
\includegraphics[width=\columnwidth]{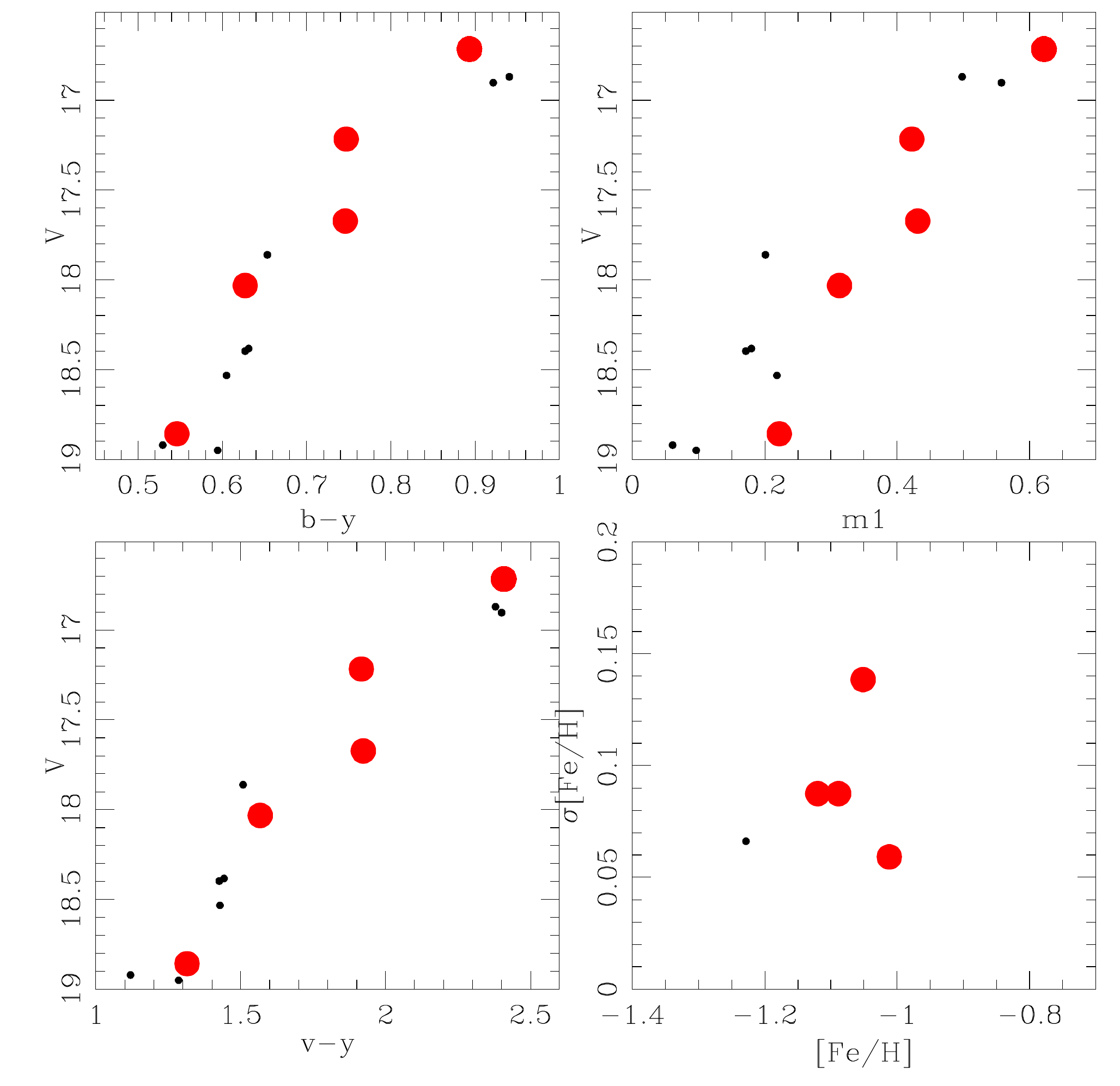}
\caption{ (continued). NGC\,2257 (left), and
ESO\,121-SC3 (riggt). }
\end{figure*}

\begin{figure}
\includegraphics[width=\columnwidth]{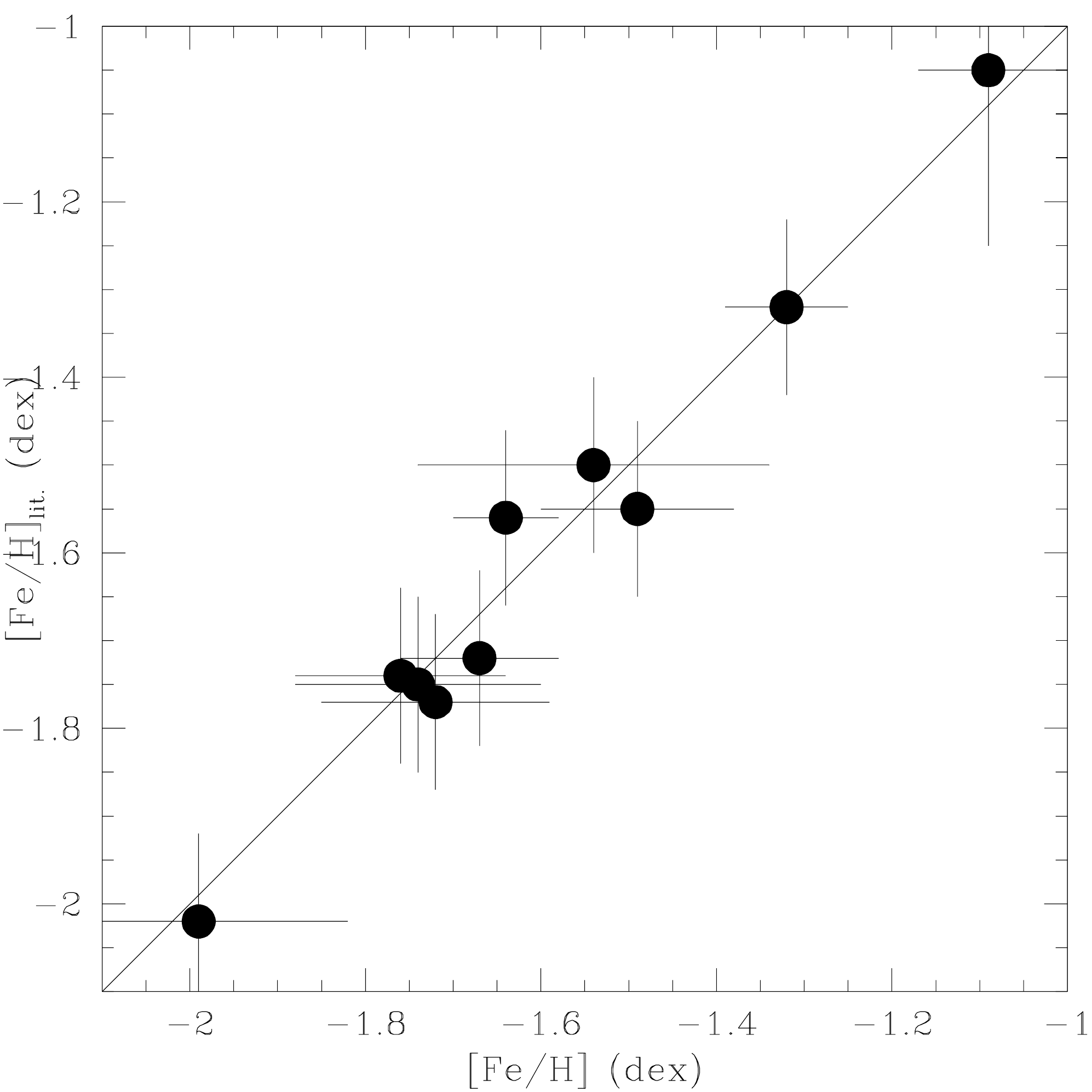}
\caption{Relationship between our [Fe/H] values and those taken
from the literature (see Table~\ref{tab:table2}).}
\label{fig3}
\end{figure}

\begin{figure}
\includegraphics[width=1\hsize]{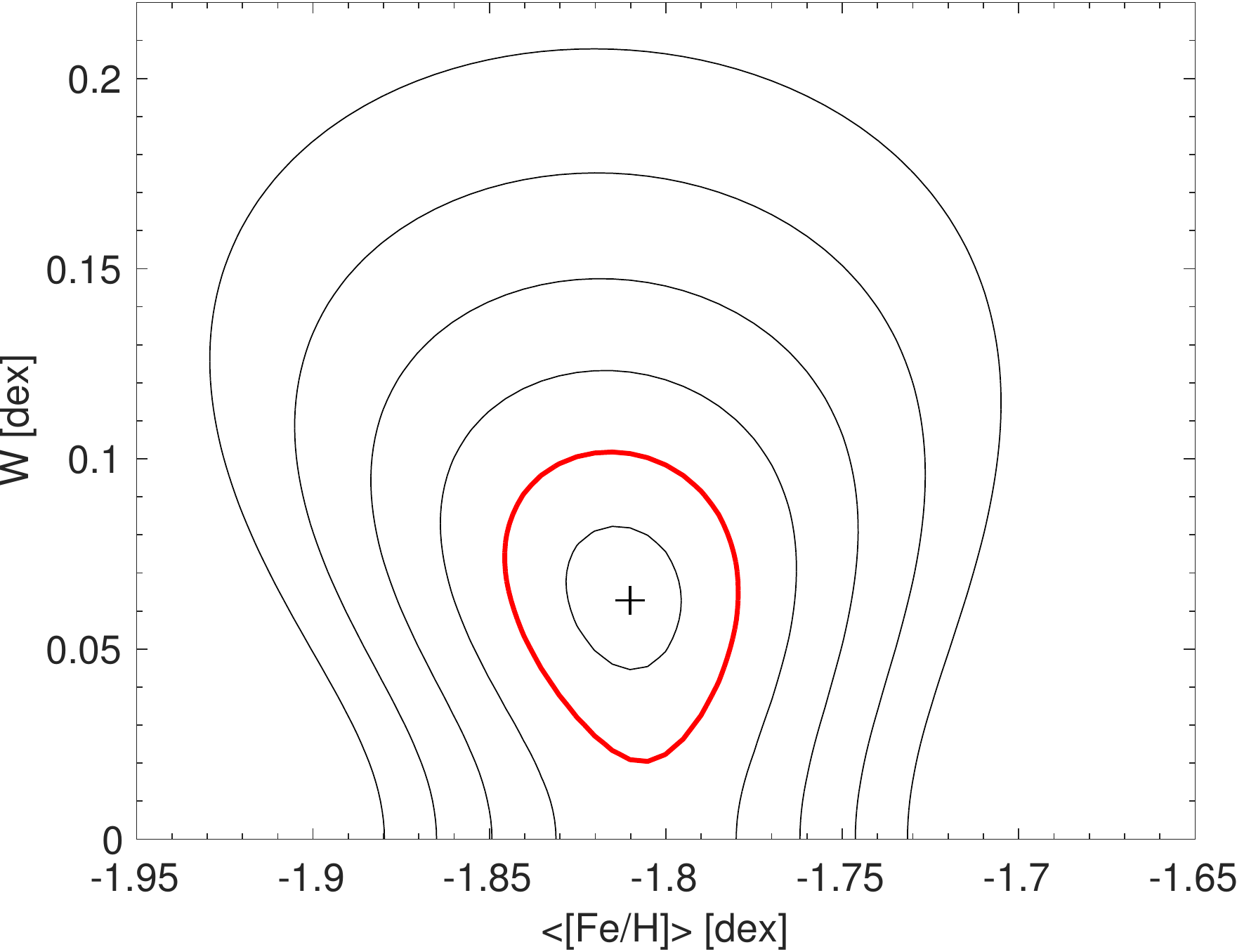}
\caption{Likelihood distribution for the mean and dispersion of NGC~1786, the only GC in our sample that shows a non-zero intrinsic dispersion.
Contours show confidence levels of 0.5--3$\sigma$ with the 68\% interval highlighted in red. }
\label{fig4}
\end{figure}

\begin{deluxetable*}{lcccccccc}
\tablecaption{Astrophysical properties of LMC GCs\label{tab:table2}}
\tablehead{\colhead{ID} & \colhead{E(B-V)$_{\rm NED}$$^a$}  & \colhead{Age} & \colhead{Ref.} & 
\colhead{[Fe/H]} &  \colhead{Ref.} & \colhead{$<$[Fe/H]$>$} & \colhead{ W} & N \\
\colhead{} & \colhead{(mag)}  & \colhead{(Gyr)} & \colhead{} & \colhead{(dex)} & & \colhead{(dex)} 
& \colhead{(dex)} & }
\startdata
NGC\,1754   & 0.07 & 12.96 & 2 &-1.50$\pm$0.10 & 4,5    &-1.54$\pm$0.02 & 0.00$\pm$0.05 & 14\\
NGC\,1786   & 0.07 & 12.30 & 7 &-1.75$\pm$0.10 &  1,4,5 &-1.80$\pm$0.03 & 0.05$\pm$0.03 & 15\\
NGC\,1835   & 0.14 & 13.37 & 2 &-1.72$\pm$0.10 &4,5     &-1.74$\pm$0.03 & 0.00$\pm$0.03 & 16\\  
NGC\,1841   & 0.14 & 12.57 & 2 &-2.02$\pm$0.10 &4       &-2.10$\pm$0.03 & 0.00$\pm$0.02 & 15\\
NGC\,1898   & 0.07 & 13.50 & 7 &-1.32$\pm$0.10 &4,5,6   &-1.32$\pm$0.02 & 0.00$\pm$0.03 & 16\\
NGC\,2005   & 0.07 & 13.77 & 2 &-1.74$\pm$0.10 &4,5,6   &-1.32$\pm$0.02 & 0.00$\pm$0.03 & 16 \\
NGC\,2019   & 0.07 & 16.20 & 2 &-1.56$\pm$0.10 &4,5,6   &-1.79$\pm$0.04 & 0.00$\pm$0.05 & 18\\ 
NGC\,2210   & 0.10 & 10.43 & 2 &-1.55$\pm$0.10 &1,2     &-1.61$\pm$0.02 & 0.00$\pm$0.04 & 9\\ 
NGC\,2257   & 0.05 & 11.54 & 2 &-1.77$\pm$0.10 & 1,4,2  &-1.80$\pm$0.05 & 0.00$\pm$0.06 & 11\\
ESO\,121-SC3& 0.04 & 8.50  & 7 &-1.05$\pm$0.20 &3       &-1.05$\pm$0.04 & 0.00$\pm$0.05 & 5\\
\enddata
\tablecomments{$^a$ Errors taken from the standard deviation of $E(B-V)$ values of the
selected stars are $<$ 0.01 mag.\\
Ref.: (1) \citet{mucciarellietal2009}; (2) \citet{wagnerkaiseretal2018}; 
(3) \citet{betal98}; (4) \citet{setal92}; (5) \citet{beetal2002}; (6) \citet{johnsonetal2006};
(7) \citet{piattietal2009}.}
\end{deluxetable*}


\acknowledgments

Based on observations obtained at the Southern Astrophysical Research (SOAR) telescope,
 which is a joint project of the Minist\'{e}rio da Ci\^{e}ncia, Tecnologia, 
Inova\c{c}\~{o}es e Comunica\c{c}\~{o}es (MCTIC) do Brasil, the U.S. National
 Optical Astronomy Observatory (NOAO), the University of North Carolina at Chapel Hill (UNC),
 and Michigan State University (MSU).
AK gratefully acknowledges support from the German Research Foundation (DFG) via 
Sonderforschungsbereich SFB 881 ('The Milky Way System', subproject A08). We thank
C.I. Johnson and A. Mucciarelli for useful comments.
We thank the referee for the thorough reading of the manuscript and
timely suggestions to improve it.


\end{document}